\documentclass[%
 reprint,
 amsmath,amssymb,
 aps,
prl,
twocolumn,
superscriptaddress,
nolongbibliography
]{revtex4-2}

\usepackage{graphicx}
\usepackage{dcolumn}
\usepackage{bm}
\usepackage{hyperref}

\usepackage{color}

\usepackage{physics}
\usepackage{mhchem}

\begin{document}

\preprint{APS/123-QED}

\title{Exploring topological spin order by inverse Hamiltonian design: A new stabilization mechanism for square skyrmion crystals}

\author{Kazuki Okigami}
\affiliation{Department of Applied Physics, The University of Tokyo, Bunkyo, Tokyo 113-8656, Japan}
\email{okigami@g.ecc.u-tokyo.ac.jp}
\author{Satoru Hayami}
\affiliation{Graduate school of Science, Hokkaido University, Sapporo 060-0810, Japan}
\email{hayami@phys.sci.hookudai.ac.jp}

\date{\today}

\begin{abstract}
We propose a framework to construct a real-space spin model based on the inverse Hamiltonian design.
The method provides an efficient way of realizing unconventional topological spin textures by optimizing the interaction parameters.
In order to demonstrate its usefulness, we show that the tuning of the long-range exchange interactions can give rise to a square skyrmion crystal even without factors that have been previously identified as prerequisites for its stabilization, such as Dzyaloshinskii-Moriya interaction, multi-spin interaction, and bond-dependent magnetic anisotropy.
Moreover, we elucidate the essence for the emergence of the square skyrmion crystal by classifying the parameter sets we get by the method.
Since the present framework by adopting machine learning techniques can be universally applied to any magnetic system irrespective of lattice structures, it serves as the efficient construction of an effective spin model and the understanding of the stabilization mechanisms for unconventional topological spin orders.
\end{abstract}

\maketitle

\textit{Introduction.---}
Magnetic materials exhibit a rich variety of fascinating spin textures at the nanoscale, driven by the interplay of competing interactions.
These complex magnetic structures not only provide insight into fundamental physics but also hold promise for next-generation devices.
Among the most intriguing are topologically non-trivial spin textures as represented by a magnetic skyrmion to have a swirling spin texture with topological stability.
These have emerged as a focal point in condensed matter physics and material science, owing to their potential applications in future computing and memory storage technologies~\cite{nagaosa2013topological,fert2013skyrmions,zhang2020skyrmion, Psaroudaki2021SkyrmionQubit}.
Characterized by their robustness against disturbances due to their non-trivial topology, skyrmions exhibit intriguing phenomena such as the topological Hall effect~\cite{Ohgushi_PhysRevB.62.R6065, Neubauer_PhysRevLett.102.186602, Hamamoto_PhysRevB.92.115417, kurumaji2019skyrmion} and the topological Nernst effect~\cite{Shiomi_PhysRevB.88.064409, Hirschberger_Nernst_prl2020}.
Since the first experimental observation of the skyrmion crystal (SkX), a periodic arrangement of skyrmions, in chiral magnets~\cite{Muhlbauer_2009skyrmion}, it has been found not only in noncentrosymmetric materials~\cite{Munzer_PhysRevB.81.041203, Yi_PhysRevB.80.054416,yu2010real, Yu2011,seki2012observation,Adams2012, Kurumaji_PhysRevLett.119.237201} but also in centrosymmetric materials~\cite{kurumaji2019skyrmion, hirschberger2019skyrmion, khanh2020nanometric,takagi2022square,yoshimochi2024multistep}.
Along with the experimental progress, theoretical studies have been extensively conducted to elucidate the stabilization mechanisms of the SkXs; the Dzyaloshinskii-Moriya (DM) interaction~\cite{dzyaloshinsky1958thermodynamic,moriya1960anisotropic} has been considered as a crucial factor for stabilizing the SkXs in noncentrosymmetric materials~\cite{rossler2006spontaneous, Yi_PhysRevB.80.054416}, while the stabilization mechanisms of the SkXs in centrosymmetric materials have been discussed in terms of the magnetic frustration of exchange interactions~\cite{Okubo_PhysRevLett.108.017206,leonov2015multiply, Lin_PhysRevB.93.064430} and the long-range interactions arising from itinerant electrons~\cite{Ozawa_PhysRevLett.118.147205, Hayami_PhysRevB.95.224424, Wang_PhysRevLett.124.207201, hayami2021topological}.
However, the stabilization mechanisms of the SkXs in centrosymmetric materials are still under debate, and the searching for a typical model to describe the SkXs is difficult owing to their complicated long-period magnetic modulations; the efficient modeling of exploring the realization of SkXs in various materials is highly desired.

Although previous research on the SkXs has succeeded in discovering several stabilization mechanisms, the conventional approach often requires tedious processes of trial-and-error in vast high-dimensional parameter spaces.
By contrast, the inverse approach, which enables us to construct an effective model from the target's physical properties, can be not only an efficient way to construct a desired model but also a powerful tool for leading us to qualitatively new physics that is difficult to reach analytically.
To construct an effective model by this approach, machine learning techniques have been widely used, such as multiple linear regression~\cite{li2020constructing, fujita2018constructing, sharma2023klm}, the Bayesian optimization~\cite{tamura2017spinspin}, and generative models~\cite{lengeling2018inverse}.
However, these techniques require data collection to some extent for training the model.
Recently, an alternative approach by using automatic differentiation has been proposed~\cite{inui2023inverse}, which calls only for the desired physical properties to construct an effective model without data collection.

In this Letter, we propose an efficient method to explore the stabilization mechanisms of topological spin orders based on the inverse Hamiltonian design~\cite{inui2023inverse}.
We construct an effective real-space spin model to stabilize the square lattice SkXs (S-SkX), which corresponds to a double-$Q$ state with a superposition of two helical states, that is akin to that observed in a centrosymmetric material \ce{GdRu2Si2}~\cite{khanh2020nanometric}.
By the aid of the machine learning techniques, we clarify that the S-SkX can be stabilized by the $xxz$-type competing exchange interaction even without the DM interaction, multi-spin interaction, or bond-dependent anisotropy, which have been thought to be necessary to stabilize the S-SkX~\cite{Hayami_PhysRevLett.121.137202, Hayami_PhysRevB.103.024439, Utesov_PhysRevB.103.064414, Wang2021meron}.
We demonstrate the emergence of the S-SkX for the obtained spin model by performing
the simulated annealing and Monte Carlo simulations.
Furthermore, by sampling the parameter sets, we find that the ratio of the interactions between a wave vector that is crucial to stabilize the triple-$Q$ state and the ordering wave vector of the S-SkX is essential in realizing the S-SkX.
These results indicate a promising framework for finding further intriguing topological spin orders that are difficult to explore by human thought.

\textit{Model.---}
Let us consider the classical spin model on a two-dimensional square lattice; the lattice constant is set as unity.
The Hamiltonian consists of the exchange interactions up to the $k_{\rm max}$-th neighbors, which is given by
\begin{equation}
    H = \sum_{k=1}^{k_{\rm max}} \sum_{\langle i,j \rangle_k} \left\{
    J^{\perp}_{k} \qty(S^{x}_i S^{x}_j
    + S^{y}_i S^{y}_j)
    + J^{zz}_{k} S^{z}_i S^{z}_j \right\},\label{eq: Hamiltonian}
\end{equation}
where $\bm{S}_i=(S_i^x, S_i^y, S_i^z)$ is a classical localized spin at site $i$ with $|\bm{S}_i|=1$, and $J^{\alpha}_{k}$ ($\alpha=\perp,zz$) is the $k$-th neighbor coupling constant for the in-plane and out-of-plane components, respectively.
$J^{\alpha}_{k} < 0 (>0)$ represents a ferromagnetic (antiferromagnetic) interaction.
The sum $\sum_{k=1}^{k_{\rm max}}$ runs over up to $k_{\rm max}$-th neighbors, and $\langle i,j \rangle_k$ denotes the $k$-th neighbor pairs.
It is noted that the model does not include the other factors, such as the DM interaction, multi-spin interaction, and bond-dependent magnetic anisotropy, which is essential for stabilizing the S-SkXs in previous studies~\cite{Hayami_PhysRevLett.121.137202, Hayami_PhysRevB.103.024439, Utesov_PhysRevB.103.064414, Wang2021meron}.

The ground state of the spin model is conjectured from the Luttinger-Tisza method~\cite{Luttinger_PhysRev.70.954} by performing the Fourier transform of the Hamiltonian in Eq.~(\ref{eq: Hamiltonian}), which is given by
\begin{align}
  H = \sum_{\bm{q}} \qty[J^{\perp}(\bm{q}) \qty(S^{x}_{\bm{q}} S^{x}_{-\bm{q}} + S^{y}_{\bm{q}} S^{y}_{-\bm{q}})
  + J^{zz}(\bm{q}) S^{z}_{\bm{q}} S^{z}_{-\bm{q}}],\label{eq: Hamiltonian wave}
\end{align}
where
  $J^{\alpha} (\bm{q}) = \sum_k \sum_{\langle i,j \rangle_k} J_k^{\alpha} e^{i \bm{q} \cdot (\bm{r}_i - \bm{r}_j)}$,
$\bm{S}_{\bm{q}}$ is the Fourier transform of $\bm{S}_i$, $\bm{q}$ is the wave vector, and $\bm{r}_i$ is the position vector of the $i$-th site.
Supposing the isotropic spin interaction, i.e., $J_k^{\perp}=J_k^{zz}$ for all $k$, the ground-state spin configuration is given by a finite-$q$ spiral state, where the ordering wave vectors are determined so as to minimize $J^{\alpha}(\bm{q})$ under the global spin-length constraint $\sum_{\bm{q}} \qty|\bm{S}_{\bm{q}}|^2 = N$, where the position of $\bm{q}$ depends on interaction parameters.
Meanwhile, the finite-$q$ spiral state can be replaced by multiple-$q$ states including the S-SkX for $J_k^{\parallel} \neq J_k^{zz}$.
Indeed, the triangular lattice SkX (T-SkX) expressed as a triple-$Q$ state can be stabilized by the Hamiltonian with the 1st-neighbor and 3rd-neighbor $xxz$-type spin interactions on a triangular lattice.
However, the S-SkX expressed as a double-$Q$ state on the square lattice has never been reported for the model in Eq.~(\ref{eq: Hamiltonian}), where the interactions are limited to up to 3rd-neighbor spins; only the T-SkX appears in the phase diagram~\cite{Lin_PhysRevB.93.064430, Wang2021meron}.

\textit{Algorithm.---}
The natural questions arise: is it possible to stabilize the S-SkX within the model in Eq.~(\ref{eq: Hamiltonian})? If yes, what are the important conditions on spin interactions?
In order to resolve these issues, we tune the parameters $J^{\perp}_k$ and $J^{zz}_k$ by adopting the machine learning techniques in the following procedure.
In the process, the cost function $C(\{J^{\perp}_k\}, \{J^{zz}_k\})$ is minimized for constructing an effective real-space spin model.
The automatic differentiation technique allows us to calculate the gradient of the cost function $\pdv{C}{J^{\alpha}_k}$ with respect to the parameters $J^{\alpha}_k$, which we then update the parameters by the gradient descent method combined with the Adam optimizer~\cite{kingma2014adam}.
The final parameters are obtained by iterating the optimization process until the cost function converges.
Hereafter, we consider the situation where the ordering wave vectors are located at $\bm{Q}_1 = (2\pi/5, 0)$, and $\bm{Q}_2 = (0,2\pi/5)$, which are related to the fourfold rotational symmetry of the square lattice.

\begin{figure}[t!]
  \begin{center}
  \includegraphics[width=0.999\hsize]{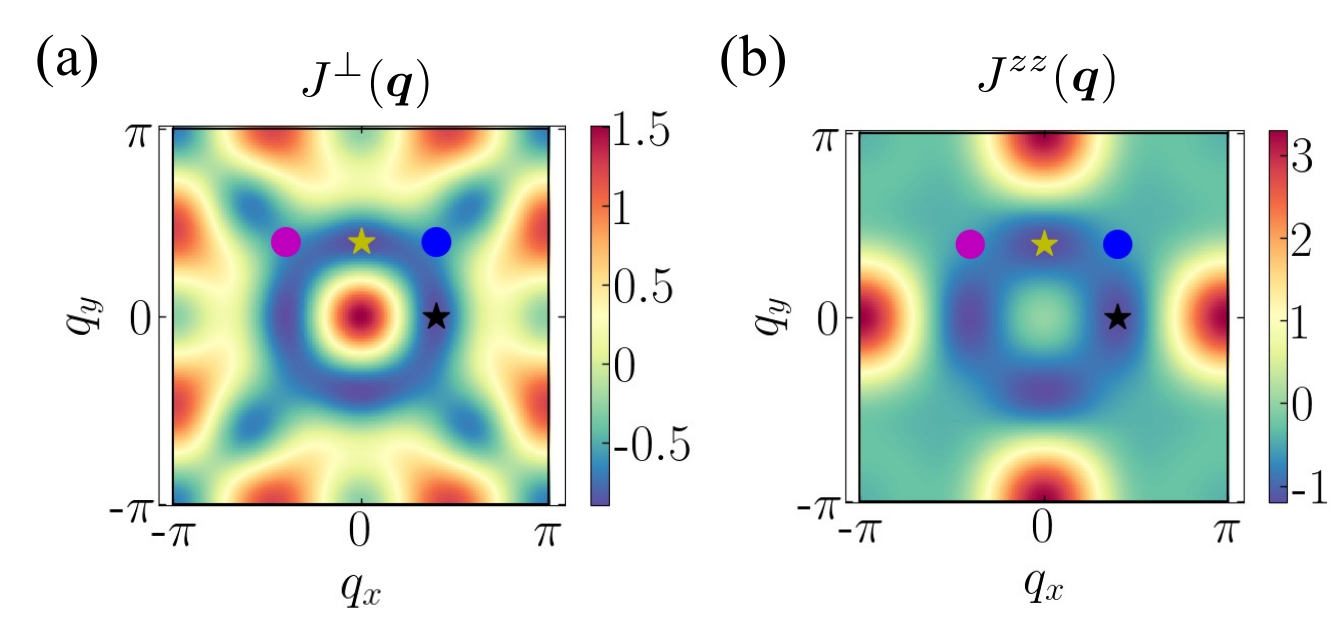}
  \caption{
  \label{fig: Jq}
  (a) The in-plane $J^{\perp} (\bm{q})$ and (b) the out-of-plane $J^{zz} (\bm{q})$ for the parameter set in Table~\ref{table: params}.
 Points marked as stars represent the ordering wave vectors $\bm{Q}_1$ (black) and $\bm{Q}_2$ (yellow), and circles represent the high-harmonic wave vectors $\bm{Q}_3$ (blue) and $\bm{Q}_4$ (pink).
  }
  \end{center}
\end{figure}

\begin{table}[t!]
  \caption{Real-space Hamiltonian parameters.}
  \label{table: params}
  \centering
  \begin{tabular}{c||c|c}
    \hline
    $k$  & $J^{\perp}$  & $J^{zz}$  \\
    \hline \hline
    1  & -0.09004769 & -0.08838843  \\
    2  & -0.06832922  & -0.3653127  \\
    3 & 0.08497451 & 0.24737181 \\
    4 & 0.11380391 & 0.03696692 \\
    5 & -0.07038448 & 0.13921292 \\
    6 & -0.00362148 & 0.01401061 \\
    7 & 0.07788382 & -0.03574327 \\
    8 & 0.07034149 & 0.02532922 \\
    \hline
  \end{tabular}
\end{table}

\begin{figure}[t!]
  \begin{center}
  \includegraphics[width=0.99\hsize]{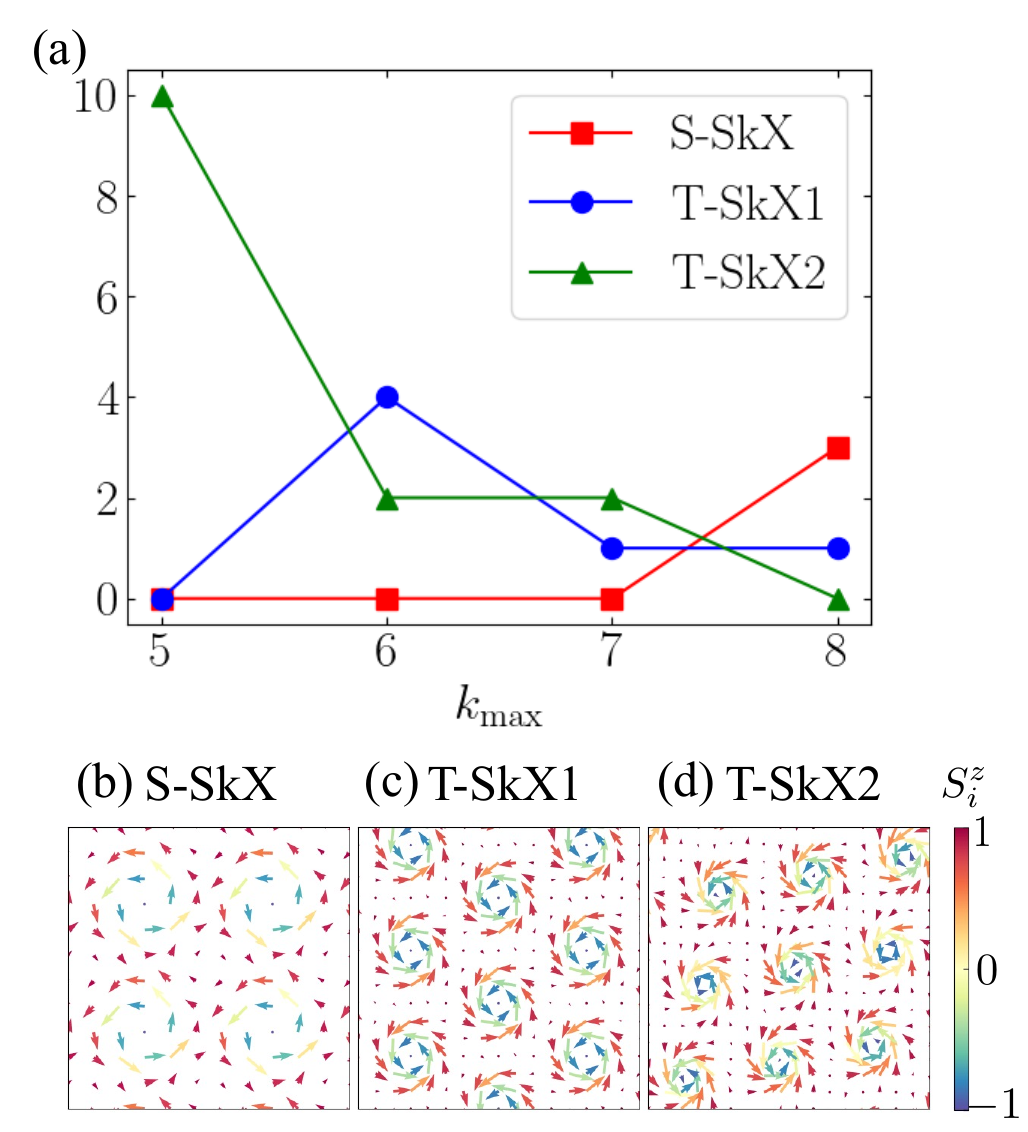}
  \caption{
  \label{fig: type of SkXs}
  (a) The ratio of the appearance of the three-type SkXs (S-SkX, T-SkX1, and T-SkX2) out of 10 independent parameter sets at each $k_{\rm max}$.
  (b)-(d) Spin configurations of the (b) S-SkX, (c) T-SkX1, and (d) T-SkX2.
  }
  \end{center}
\end{figure}

The specific expression of the cost function is set as
\begin{align}
  C (\{J^{\perp}_k\}, \{J^{zz}_k\}) &= \sum_{\bm{q} = \bm{Q}_1, \bm{Q}_3
  ,\bm{0}}
  \qty(J^{\alpha}(\bm{q}) - \tilde{X}^{\alpha}(\bm{q}))^2 \nonumber\\
  &+ \sum_{\alpha = \perp,zz} \sum_{\bm{q}\neq \bm{Q}_1, \bm{Q}_2}
  \qty[\mathrm{ReLu} \qty(J^{\alpha}(\bm{q}) - J^{\alpha} (\bm{Q}_1))] \nonumber\\
  &+ \left. \sum_{\alpha = \perp,zz} \qty(\pdv{J^{\alpha}(\bm{q})}{\bm{q}})^2 \right|_{\bm{q} = \bm{Q}_1},\label{eq: cost function}
\end{align}
where $\tilde{X}^{\alpha}(\bm{q})$ represents the target $J^{\alpha}(\bm{q})$, where the tilde symbol denotes the given parameters to obtain the desired magnetic structures. $\mathrm{ReLu} (x) = \mathrm{max} (0,x)$ is the rectified linear unit (ReLu) function.
We disregard $\bm{Q}_2$ in the cost function since we keep the fourfold rotational symmetry of the square lattice during the optimization process.
As all terms in the cost function are always non-negative, the cost function is minimized when the model exhibits the target $J^{\alpha}(\bm{q})$ and has global minima at $\bm{Q}_1$ and $\bm{Q}_2$.

The first term in the cost function is for achieving the target $J^{\alpha}(\bm{q})$.
We set $\tilde{X}^{\perp} (\bm{Q}_1) = -0.9$, $\tilde{X}^{zz} (\bm{Q}_1) = -1.2$, $\tilde{X}^{\perp} (\bm{Q}_{3}) = \tilde{X}^{zz} (\bm{Q}_{3}) = -0.6$, and $\tilde{X}^{zz} (\bm{0}) = 0$, where $\bm{Q}_3$ corresponds to the high-harmonic wave vector of $\bm{Q}_1$ and $\bm{Q}_2$, i.e., $\bm{Q}_3=\bm{Q}_1 + \bm{Q}_2 = (2\pi/5, 2\pi/5)$.
Those values are determined to be optimal for the model to prefer the S-SkX. $\tilde{X}^{\perp}(\bm{Q}_1)$ and $\tilde{X}^{zz}(\bm{Q}_1)$ are set to be negative and minima of $J^{\perp}(\bm{q})$ and $J^{zz}(\bm{q})$ under the condition of the easy-axis anisotropy $\tilde{X}^{\perp}(\bm{Q}_1) > \tilde{X}^{zz}(\bm{Q}_1)$.
In addition to this, it has a non-negligible contribution by the higher harmonics $\bm{Q}_3$ and $\bm{Q}_4=-\bm{Q}_1 + \bm{Q}_2$.
This significance of the higher-harmonic component is unique for stabilizing the S-SkX~\cite{hayami2023widely}.
This is understood from the fact that the sum of {$\bm{Q}_1$, $\bm{Q}_2$, and} $\bm{Q}_3$ (or $\bm{Q}_4$) leads to $\bm{Q}_1+\bm{Q}_2-\bm{Q}_3 = \bm{0}$ (or $\bm{Q}_1-\bm{Q}_2+\bm{Q}_4 = \bm{0}$), which indicates the effective coupling in the form of $(\bm{S}_{\bm{0}}\cdot \bm{S}_{\bm{Q}_1})(\bm{S}_{\bm{Q}_2} \cdot \bm{S}_{-\bm{Q}_3})$ favoring the multiple-$Q$ state, as found in the T-SkX with $\bm{Q}'_1+\bm{Q}'_2+\bm{Q}'_3=\bm{0}$~\cite{Muhlbauer_2009skyrmion} ($\bm{Q}'_1$, $\bm{Q}'_2$, and $\bm{Q}'_3$ are regarded as the triple-$Q$ ordering wave vectors, whose relative angles are $120^{\circ}$).

The second and third terms are for ensuring that $J^{\alpha}(\bm{q})$ takes the minimum values at $\bm{Q}_1$ and $\bm{Q}_2$.
The former is the function for penalizing the model when $J^{\alpha}(\bm{q}\neq \bm{Q}_1, \bm{Q}_2)$ is smaller than $J^{\alpha}(\bm{Q}_1)$.
The latter is the square of the derivatives of $J^{\alpha}(\bm{q})$ with respect to $q_x$ and $q_y$ at $\bm{Q}_1$.
This term penalizes the model when the derivatives of $J^{\alpha}(\bm{q})$ at $\bm{Q}_1$ are not zero, which ensures the peak structures at $\bm{Q}_1$ and $\bm{Q}_2$.
We show the contour plot of $J^{\perp}(\bm{q})$ [$J^{zz}(\bm{q})$] in Fig.~\ref{fig: Jq}(a) [Fig.~\ref{fig: Jq} (b)] for the parameter set in Table~\ref{table: params}, which is obtained by the above procedure; both $J^{\perp}(\bm{q})$ and $J^{zz}(\bm{q})$ smoothly change in the momentum space while satisfying the model parameter conditions, as stated above.

\textit{Inverse Hamiltonian design.---}
As the inverse problem regarding the cost function in Eq.~(\ref{eq: cost function}) is underdetermined, that is to say, there are multiple solutions that minimize the cost function due to the lack of equations against the number of parameters, we examine 10 independent parameter sets and classify the types of SkXs for $k_{\rm max}=5$--$8$.
The cost function is minimized to the order of $10^{-8}$--$10^{-11}$ for all the parameter sets, where all the parameters are shown in the supplemental material~\cite{suppl}.
Figure 2(a) shows the ratio of the appearance of SkXs for the spin model with the $k_{\rm max}$-th neighbor interactions.
We obtain three types of the SkX, which are classified into the S-SkX with the double-$Q$ peaks at $\bm{Q}_1$ and $\bm{Q}_2$ and two T-SkXs with the triple-$Q$ peaks at $\bm{Q}'_1$, $\bm{Q}'_2$, and $\bm{Q}'_3$ satisfying $\bm{Q}'_1+\bm{Q}'_2+\bm{Q}'_3=\bm{0}$ in the spin structure factor~\cite{3Qnote}
The latter T-SkX is further classified by the position of the triple-$Q$ ordering wave vectors: one is the case when $\bm{Q}'_1=\bm{Q}_1$ while the other is the case when $\bm{Q}'_1$ slightly deviates from $\bm{Q}_1$; we refer to the former as T-SkX1 and the latter as T-SkX2.
The snapshots of the S-SkX, T-SkX1, and T-SkX2 are shown in Figs.~\ref{fig: type of SkXs}(b), \ref{fig: type of SkXs}(c) and \ref{fig: type of SkXs}(d), respectively~\cite{suppl}.

As shown in Fig.~\ref{fig: type of SkXs}(a), the S-SkX is stabilized only for $k_{\rm max}=8$, while the T-SkX1 and/or T-SkX2 are stabilized for smaller $k_{\rm max}$.
For $k_{\rm max}=8$, the S-SkX is realized for three out of ten parameter sets, which supports a relatively high probability of getting the optimal parameters.
These results clearly indicate that the frustrated long-range exchange interaction tends to stabilize the S-SkX even without the multi-spin interaction or the bond-dependent anisotropy.
Since it is usually cumbersome to tune the interaction parameters by the conventional analytical approach, our method based on the inverse Hamiltonian design provides a powerful method to derive unconventional topological spin orders that have never been clarified by a simple spin model.

\begin{figure}[t!]
  \begin{center}
  \includegraphics[width=0.99\hsize]{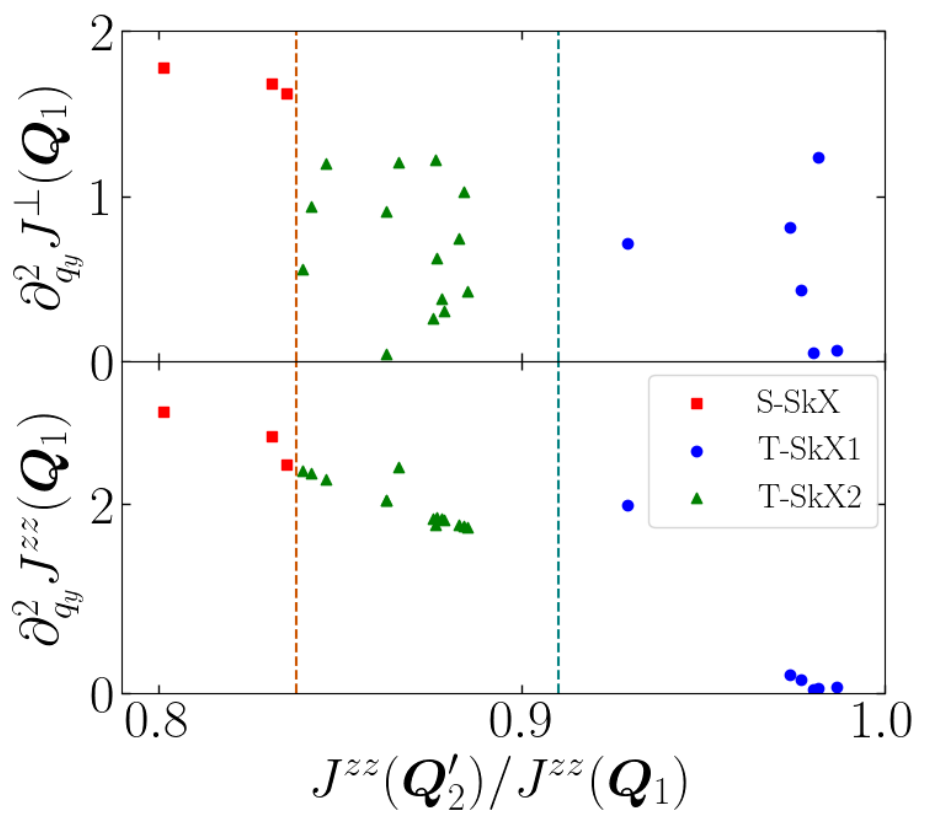}
  \caption{
  \label{fig: parameter classification}
  Classification of the parameter sets that stabilize each type of SkX.
  }
  \end{center}
\end{figure}

On the other hand, the emergence of the different types of the SkXs in Fig.~\ref{fig: type of SkXs}(a) seems to be surprising, since the dominant interactions $J^{\perp}(\bm{Q}_\nu)$ and $J^{zz}(\bm{Q}_\nu)$ are common for all the independent parameter sets irrespective of $k_{\rm max}$.
To clarify the origin of the different SkX tendencies for different parameter sets, we try to extract the essence by further analyses.
As a result, we find two essences to induce the S-SkX rather than the T-SkX.
One significant factor is the ratio of $J^{zz} (\bm{Q}'_2)$ to $J^{zz} (\bm{Q}_1)$, where $\bm{Q}'_2=R(2\pi/3) \bm{Q}_1$, where $R(\theta)$ is the rotation matrix by $\theta$.
Since the interaction at $\bm{Q}'_2$ contributes to only the T-SkX1 rather than the S-SkX, the small ratio tends to favor the S-SkX, as shown in Fig.~\ref{fig: parameter classification}.
In other words, the T-SkX1 is favored when $J^{zz} (\bm{Q}'_2)$ is comparable to $J^{zz} (\bm{Q}_1)$.

The second is the second derivatives of $J^{\perp} (\bm{q})$ and $J^{zz} (\bm{q})$ with regard to $q_y$ at $\bm{Q}_1$.
This is because the second derivatives represent the steepness of the peaks of $J^{\alpha} (\bm{q})$ at $\bm{Q}_1$; the larger the second derivatives are, the harder it is to displace the ordering wave vector from $\bm{Q}_1$.
Thus, the larger second derivatives assist the stabilization of the S-SkX compared to the T-SkX2.

\textit{Phase diagram.---}
Once the effective spin model is constructed, one can calculate the magnetic-field---temperature phase diagram by the Monte Carlo (MC) simulations and simulated annealing in the following steps.
First, we perform the MC simulations combined with the parallel tempering method~\cite{hukushima1996exchange} to avoid being trapped in local minima.
Each simulation is performed for $10^6-10^7$ MC sweeps to reach thermal equilibrium, and then the data are collected by calculating the average of the physical variables for the next $10^6-10^7$ MC sweeps.
Hereafter, we show the results for the system size $L=60$ with the total number of spins $N=60^2$ under the periodic boundary condition, but the results for the larger system sizes $L=80$ and $100$ are qualitatively the same as shown in the supplemental material~\cite{suppl}.
After that, we perform the simulated annealing starting from spin configurations at the minimum temperature of the MC simulations $T=0.05$ to the final temperature $T=0.001$ for $10^6$ MC sweeps.

\begin{figure}[t!]
\begin{center}
\includegraphics[width=0.99\hsize]{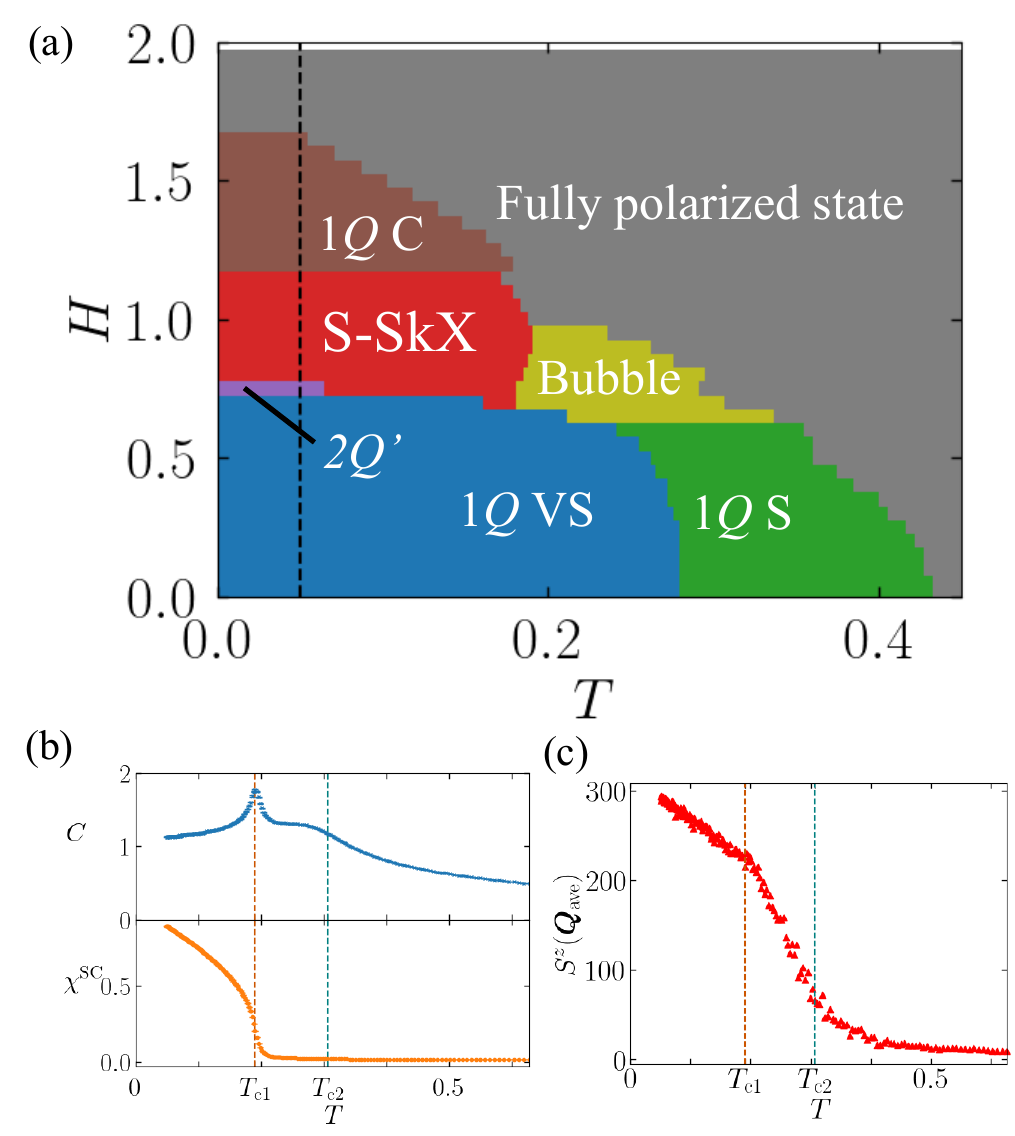}
\caption{
\label{fig: phase diagram}
(a)The $H$-$T$ phase diagram of the model with the parameter set in Table~\ref{table: params}.
(b) (upper panel) Specific heat $C$, and (lower panel) net scalar spin chirality $\chi^{\rm SC}$ at $H=0.9$.
(c) $T$ dependence of the $\bm{Q}_1$ and $z$-spin components of the spin structure factor.
}
\end{center}
\end{figure}

The obtained $H$-$T$ phase diagram is presented in Fig.~\ref{fig: phase diagram}(a), where we adopt the parameter set in Table~\ref{table: params}.
The effect of the magnetic field $H$ is introduced as $-H \sum_i S^z_i$.
The phase boundaries are determined by calculating the specific heat $C$, the net scalar spin chirality $\chi^{\rm SC}$, and the spin structure factors $S^{\alpha} (\bm{q})$; $\chi^{\rm SC} = \frac{1}{N} \sum_{\langle i,j,k \rangle} \bm{S}_i \cdot (\bm{S}_j \times \bm{S}_k)$, where $\langle i,j,k \rangle$ denotes the triangular plaquette, and $S^{\alpha} (\bm{q}) = \frac{1}{N} \sum_{i,j} \expval{S^{\alpha}_i S^{\alpha}_j} e^{i \bm{q} \cdot (\bm{r}_i - \bm{r}_j)}$.
In the lower $H$ region, the ground state is the $1Q$ vertical spiral state ($1Q$ VS), which is characterized by the ordering wave vector $\bm{Q}_1$ or $\bm{Q}_2$.
By increasing $H$, the system undergoes a phase transition to a $2Q'$ state with the unequal amplitudes of $\bm{Q}_1$ and $\bm{Q}_2$, then to the S-SkX state, $1Q$ conical state ($1Q$ C), and the fully polarized state.

In Figs.~\ref{fig: phase diagram}(b) and \ref{fig: phase diagram}(c), $T$ dependence of $C$, $\chi^{\rm SC}$, and the $z$-component of the spin structure factor $S^{z} (\bm{Q}_{\rm ave})$ at $H=0.9$ are shown, respectively, where $S^{z} (\bm{Q}_{\rm ave})$ represents the averaged spin structure factor over $\bm{Q}_1$ and $\bm{Q}_2$.
The specific heat $C$ exhibits a peak at $T_{\rm c1} \simeq  0.19$ where $\chi^{\rm SC}$ is developed, which means the appearance of the S-SkX.
Meanwhile, $S^{z} (\bm{Q}_{\rm ave})$ slowly increases with decreasing $T$ from a higher temperature than $T_{\rm c1}$, which indicates the appearance of the Bubble phase for $T_{\rm c1} \lesssim T \lesssim T_{\rm c2}$ with $T_{\rm c2} \simeq 0.30$~\cite{Hayami_PhysRevB.108.024426}.
As in the case of $H=0.9$, only the $z$-component of spins orders at higher temperatures irrespective of $H$, thus the $1Q$ sinusoidal state ($1Q$ S) appears in the lower $H$ region.

\textit{Conclusion.---}
We have shown that the inverse approach can find an unexpected physics that is difficult to reach analytically.
We have constructed an effective real-space spin model by the inverse Hamiltonian design method by targeting the stabilization of the S-SkX.
By performing the MC simulations and simulated annealing, we have demonstrated that the S-SkX is stabilized from zero to finite temperatures even without multiple-spin interaction or bond-dependent anisotropy.

Our findings suggest that the discovery of a different stabilization mechanism for the S-SkX represents a significant advancement in our understanding of conventional SkX stabilization mechanisms that have been established in previous studies~\cite{Hayami_PhysRevB.103.024439, Utesov_PhysRevB.103.064414, Wang2021meron}.
As an advantage, this method circumvents the laborious process of trial and error in an extensive high-dimensional parameter space that is inherent to the conventional approach.
Moreover, this method is applicable to any type of magnetic structure, thereby it paves the way for the exploration of unveiled physics rooted in diverse magnetic structures.

\textit{Acknowledgments.---}
We thank Koji Inui for fruitful discussions on the algorithm.
This work was supported by JST SPRING, Grant Number JPMJSP2108.
This was also supported by JSPS KAKENHI Grants Numbers JP21H01037, JP22H04468, JP22H00101, JP22H01183, JP23H04869, JP23K03288, JP23K20827, and by JST PRESTO (JPMJPR20L8) and JST CREST (JPMJCR23O4).
KO would like to acknowledge the support from the Motizuki Fund of Yukawa Memorial Foundation.

\bibliography{ref}

\begin{thebibliography}{49}
\bibitem [{3Qn()}]{3Qnote}%
  \BibitemOpen
  \href@noop {} {}\bibinfo {note} {Since we work on finite-size systems, $\bm{Q}_2'$ and $\bm{Q}_3'$ are not exactly related with $\bm{Q}_1'$ by the $120^\circ$ rotation symmetry. The finite-size effect is expected to be small, but it may cause the slight deviation of the results from the ideal three-$\bm{Q}$ state.}\BibitemShut {Stop}%
\end{thebibliography}

\end{document}